\documentclass[doublecol, figures]{epl2} 
\usepackage{blindtext}
\usepackage{subcaption}
\usepackage{amsmath}
\usepackage{color} 
\usepackage[table]{xcolor}
\usepackage{graphicx}

\title{Interplay of synchronization and cortical input in models of brain networks}
\shorttitle{Interplay of synchronization and cortical input in models of brain networks} 

\author{Jakub Sawicki\inst{1,2,3,4*} \and Eckehard Sch{\"o}ll\inst{1,4,5}}
\shortauthor{J. Sawicki \etal}

\institute{                    
  \inst{1} Potsdam Institute for Climate Impact Research - Telegrafenberg A 31, 14473 Potsdam, Germany \\
  \inst{2} Institut f{\"u}r Musikp{\"a}dagogik, Universit{\"a}t der K{\"u}nste Berlin - Hardenbergstra\ss{}e 33, 10623 Berlin, Germany \\
  \inst{3} Fachhochschule Nordwestschweiz FHNW - Leonhardsstrasse 6, 4009 Basel, Switzerland \\
  \inst{4} Bernstein Center for Computational Neuroscience Berlin, Humboldt-Universit{\"a}t - Philippstra{\ss}e 13, 10115 Berlin, Germany \\ 
  \inst{5} Institut f{\"u}r Theoretische Physik, Technische Universit{\"a}t Berlin - Hardenbergstra\ss{}e 36, 10623 Berlin, Germany \\
  * zergon@gmx.net
}


\abstract{ It is well known that synchronization patterns and coherence have a major role in the functioning of brain networks, both in pathological and in healthy states. In particular, in the perception of sound, one can observe an increase in coherence between the global dynamics in the network and the auditory input. In this perspective article, we show that synchronization scenarios are determined by a fine interplay between network topology, the location of the input, and frequencies of these cortical input signals. To this end, we analyze the influence of an external stimulation in a network of FitzHugh-Nagumo oscillators with empirically measured structural connectivity, and discuss different areas of cortical stimulation, including the auditory cortex. 
}

\begin{document}

\maketitle

\section{Introduction}

Synchronization is a common phenomenon occurring in 
coupled systems~\cite{PIK01,BOC18}. 
Synchronization phenomena play a particularly important role in the context of brain dynamics. 
In the human brain, synchronization of neurons is essential~\cite{SIN18} for normal physiological functioning, e.g., in the context of cognition and learning, but it is also closely related to pathological conditions such as Parkinson's disease \cite{TAS98} or epileptic seizures, which are the cardinal symptom of epilepsy. Here, synchronized dynamics plays an important role \cite{GER20}, in which synchronization of one part of the brain has dangerous consequences for the affected person. 

On the other hand, synchronization is also used  as a paradigm to explain brain processes that serve the development of syntax and its perception \cite{KOE13,LAR15a,BAD20,SAW23a}. In general, synchronization theory is of great importance for the analysis and understanding of musical acoustics and music psychology \cite{JOR94,BAD13,SAW18a,HOU20a,SHA20}. While the neurophysiological processes involved in listening to music continue to be explored, it is believed that some degree of synchrony can be observed in listening to music, both associated with attention and developing expectations. Event-related potentials measured by electroencephalography (EEG) from participants while listening to music show synchronized dynamics between different brain regions~\cite{HAR20a}. These studies suggest that the increase in synchronization represents large-form musical perception, in which the frequency of the sound evokes synchronization patterns in the brain~\cite{BAD13,SAW21a}. Moreover, areas of the whole brain have been observed to be involved in neural dynamics during perception \cite{BAD20}. Recent studies report an increase of coherence between the global dynamics and the input signal induced by a specific music song using measured electroencephalogram (EEG) data \cite{SAW22}. Therefore, it is relevant to investigate the influence of sound on empirical brain networks in dependence of the auditory input region. We model the spiking dynamics of neurons using the FitzHugh-Nagumo (FHN) system, which is a paradigmatic model for neural dynamics~\cite{BAS18}, and investigate possible partial synchronization patterns induced by an external sound source connected to the auditory cortex of the human brain. It is well known that an important feature of musical sound perception is sound fusion\cite{SCH18k}. Although sound in general has a rich overtone spectrum, subjects perceive only one pitch, which is a fusion of all partials of the spectrum. 

In this perspective article we focus on an external sound source with fixed amplitude and a single frequency and neglect the complexity of music and its different effects in different frequency bands within brain oscillations. In the context of this work, we restrict ourselves to a minimal model without node-specific behavior to demonstrate the effects of a periodic perturbation~\cite{SAW21a}. More complex input using the transformation of sound into electrical neural signals via an elaborate cochlea model has been reported elsewhere~\cite{SAW22}. Using networks of coupled neural oscillators, one can simulate synchronization phenomena observed in the human brain. The FHN model describes the nonlinear dynamics of individual neurons or whole brain areas by a fast excitatory and a slow inhibitory variable. The coupling between different neurons or different areas of the brain is mediated by a coupling matrix, which may be mathematically constructed using standard procedures from network science, or taken from empirical structural brain connectivities of human subjects measured, e.g, by diffusion-weighted magnetic resonance imaging (MRI)~\cite{SAN15a,SKO22}. The role of distant-dependent transmission time delays in large-scale brain synchronization has been stressed in \cite{PET19}.
Here, we focus on the interplay of synchronization and cortical input into different brain areas, characterized by the 90 regions of the human brain according to the Automated Anatomic Labeling (AAL) atlas. 

\section{Synchronization scenarios}

Synchronization may not only occur as complete or isochronous (zero-lag) synchronization, but also as partial synchronization of parts of the system~\cite{SCH21}. Also, frequency synchronization (in which only the frequencies but not the phases are the same) should be distinguished from phase synchronization (in which the phases are also synchronized). Partial synchronization includes cluster or group synchronization (in which within each cluster all the elements are synchronized, but there is no synchrony between the clusters), and solitary states (in which isolated nodes form single-node clusters), and chimera states (spontaneously symmetry-breaking states with partially synchronized (spatially coherent) and partially desynchronized (spatially incoherent) dynamics~\cite{KUR02a,ABR04,SAW20,ZAK20}).

\section{Model}

\begin{figure}[tp]
    \begin{center}
\includegraphics[width = \linewidth]{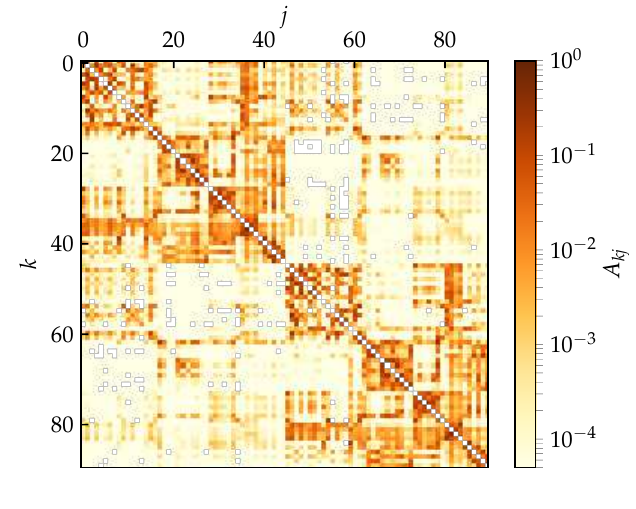}
    \caption{(color online) Model for the brain structure: Weighted adjacency matrix $A_{kj}$ of the averaged empirical structural brain network derived from twenty healthy human subjects by averaging over the coupling between two brain regions $k$ and $j$. 
    After \cite{GER20,SKO22}.}
    \label{fig.1}
    \end{center}
\end{figure}

We consider an empirical structural brain network (Fig.\,\ref{fig.1}) where every brain area is modeled by a single FitzHugh-Nagumo (FHN) oscillator\cite{FIT61,NAG62,CHE05e,CHE07a,BAS18}.

The weighted adjacency matrix $\mathbf{A} = \{A_{kj}\}$ of size $90 \times 90$, with node indices $k \in \{1,2,...,90\}$ was obtained from averaged diffusion-weighted magnetic resonance imaging data\cite{SKO22} divided into 90 predefined regions according to the Automated Anatomical Labeling (AAL) atlas \cite{TZO02} as described in Table \ref{tab:AAL}, for previous use in simulations see \cite{GER20,SCH21,SAW21a,SAW22}.
Each node of the network corresponds to a brain region. Note that in contrast to the original AAL indexing, where sequential indices correspond to homologous brain regions, the indices in Fig.\,\ref{fig.1} are rearranged such that $k \in N_L = \{1, 2, ... ,45\}$ corresponds to left and $k \in N_R = \{46, ... ,90\}$ to the right hemisphere. Thereby the hemispheric structure of the brain, i.e., stronger intra-hemispheric coupling compared to inter-hemispheric coupling, is highlighted (Fig.\,\ref{fig.1}). 

The auditory cortex is the part of the temporal lobe that processes auditory information in humans. It is a part of the auditory system, performing basic and higher functions in hearing and is located bilaterally, roughly at the upper sides of the temporal lobes, i.e., corresponding to the AAL indexing $k = 41,86$ (temporal sup L/R). The auditory cortex takes part in the spectrotemporal analysis of the inputs passed on from the ear.

The FHN dynamics of the network with external stimulus reads:

\begin{subequations}
\begin{align}
\epsilon \dot{u}_k = &u_k - \frac{u_k^3}{3} - v_k \nonumber \\
                    &+ \sigma \sum_{j \in N_\text{H}} A_{kj} \left[ B_{uu}(u_j - u_k) + B_{uv}(v_j - v_k) \right]  \\
                    &+ \varsigma \sum_{j \notin N_\text{H}} A_{kj} \left[ B_{uu}(u_j - u_k) + B_{uv}(v_j - v_k) \right], \nonumber \\
                     & +C_k^I\gamma \cos \omega t \nonumber\\ \nonumber\\
\dot{v}_k = &u_k + a \nonumber \\
 & + \sigma \sum_{j \in N_\text{H}} A_{kj} \left[ B_{vu}(u_j - u_k) + B_{vv}(v_j - v_k) \right]  \\
 & + \varsigma \sum_{j \notin N_\text{H} } A_{kj} \left[ B_{vu}(u_j - u_k) + B_{vv}(v_j - v_k) \right], \nonumber
\end{align}
\label{eq.1}
\end{subequations}
with $k \in N_\text{H}$ where $N_\text{H}$ denotes either the set of nodes $k$ belonging to the left ($N_L$) or the right ($N_R$) hemisphere, and $\epsilon = 0.05$ describes the timescale separation between the fast activator variable (neuron membrane potential) $u$ and the slow inhibitor (recovery variable) $v$ \cite{FIT61}. Depending on the threshold parameter $a$, the FHN model may exhibit excitable behavior ($\left| a \right| \ge 1$); or self-sustained oscillations ($\left| a \right| < 1$). For all nodes we use the oscillatory regime in order to obtain clear insight into the topological effect of different locations of the input, and we fix $a=0.5$ sufficiently far from the Hopf bifurcation point. The external input stimulus is modeled by a trigonometric function with frequency $\omega$ and amplitude $\gamma$ and is applied to a specific pair of cortical regions $k=I_0$ and $k=I_0+45$, where the index $I=I_0$ denotes the stimulated area. For instance, $k=41$ and $k=86$ are associated with the auditory cortex, i.e. $C_k^I=1$ if $k=41$ or $k=86$ and zero otherwise. The intra-hemispheric coupling between the single regions is given by the coupling strength $\sigma$, and the inter-hemispheric coupling is given by $\varsigma$. As we are looking for partial synchronization patterns we fix $\sigma = 0.7$ and $\varsigma=0.15$ similar to numerical studies of synchronization phenomena during unihemispheric sleep \cite{SCH21} and epileptic seizures \cite{GER20} where partial synchronization patterns have been observed. The interaction scheme between nodes is characterized by a rotational coupling matrix:

\begin{equation}
\mathbf{B} = 
\begin{pmatrix}
B_{uu} & B_{uv} \\
B_{vu} & B_{vv}
\end{pmatrix}
=
\begin{pmatrix}
\text{cos}\phi & \text{sin}\phi \\
-\text{sin}\phi & \text{cos}\phi
\end{pmatrix},
\end{equation}
with coupling phase $\phi = \frac{\pi}{2} - 0.1$, causing dominant activator-inhibitor cross-coupling~\cite{OME13}, i.e., the prevalence of excitatory vs. inhibitory couplings. Also in the modeling of epileptic-seizure-related synchronization phenomena \cite{GER20}, where a part of the brain synchronizes, it turned out that such a cross-coupling is important. 
Mathematically, this means that signals of other neuronal areas are coupled via a coupling phase, which introduces a phase lag or time delay. The subtle interplay of excitatory and inhibitory interaction enables intermittent periods of either high or low synchronization. This is typical of the critical state at the edge of different dynamical regimes in which the brain operates~\cite{MAS15a,WIL19,SHI22}.

\section{Methods}

The dynamics can be characterized by the mean phase velocity ${\Omega_k = 2\pi M_k/\Delta T}$ of each node $k$, where $\Delta T$ denotes the time interval during which $M$ complete rotations are realized. Throughout the paper we use $\Delta T = 10\,000$. For all simulations we use initial conditions randomly distributed on the circle $u_k^2+v_k^2=4$. In case of an uncoupled system ($\sigma=\varsigma=0$), the mean phase velocity (or natural frequency) of each node is $\Omega_k = \omega_{\text{FHN}} \approx 2.6$. Furthermore we introduce measures that characterize the degree of synchronization. First, the spatially averaged mean phase velocity is: 
\begin{equation}
{\overline\Omega = \frac{1}{N}\sum_{k =1}^N \Omega_k},    
\end{equation}
Second, the Kuramoto order parameter:
\begin{equation}
{R(t) = \frac{1}{N} \left| \sum_{k =1}^N \text{exp}[i \theta_k(t)]\right|},
\end{equation}
is calculated by means of an abstract dynamical phase $\theta_k$ that can be obtained from the standard geometric phase ${\tilde{\phi}_k(t) = \text{arctan}(v_k/u_k)}$ by a transformation which yields constant phase velocity $\dot{\theta}_k$. For an uncoupled FHN oscillator the function $t(\tilde{\phi}_k)$ is calculated numerically, assigning a value of time $0<t(\tilde{\phi}_k)<T$ for every value of the geometric phase, where $T$ is the oscillation period. The dynamical phase is then defined as $\theta_k=2 \pi t(\tilde{\phi}_k)/T$, which yields $\dot{\theta}_k = \text{const}$. Thereby temporal fluctuations of the order parameter $R$ caused by the FHN model's slow-fast time scales are suppressed and a change in $R$ indeed reflects a change in the degree of synchronization. The Kuramoto order parameter may vary between 0 and 1, where $R=1$ corresponds to complete phase synchronization, and small values characterize spatially desynchronized states. Additionally, we calculate the temporal mean of the Kuramoto order parameter
\begin{equation}
\langle R(t) \rangle = \frac{1}{\Delta T}\int_0^{\Delta T} R(t) \,\mathrm{d} t
\end{equation}
to estimate the general dynamical behavior of the system over time. Similarly, the temporal mean $\langle \Omega(t) \rangle $ of the collective frequency $\Omega$ of the mean field \cite{PET13}, defined by
\begin{equation}
\Omega(t)\equiv \dot{\psi}(t), \quad     R(t) e^{i \psi(t)} = \frac{1}{N} \sum_{k =1}^N \text{exp}[i \theta_k(t)]
\end{equation}
can be considered, and compared with the spatially averaged mean phase velocity.

\section{Role of auditory cortex}

\begin{figure}
\centering
\includegraphics[width = 1.0\linewidth]{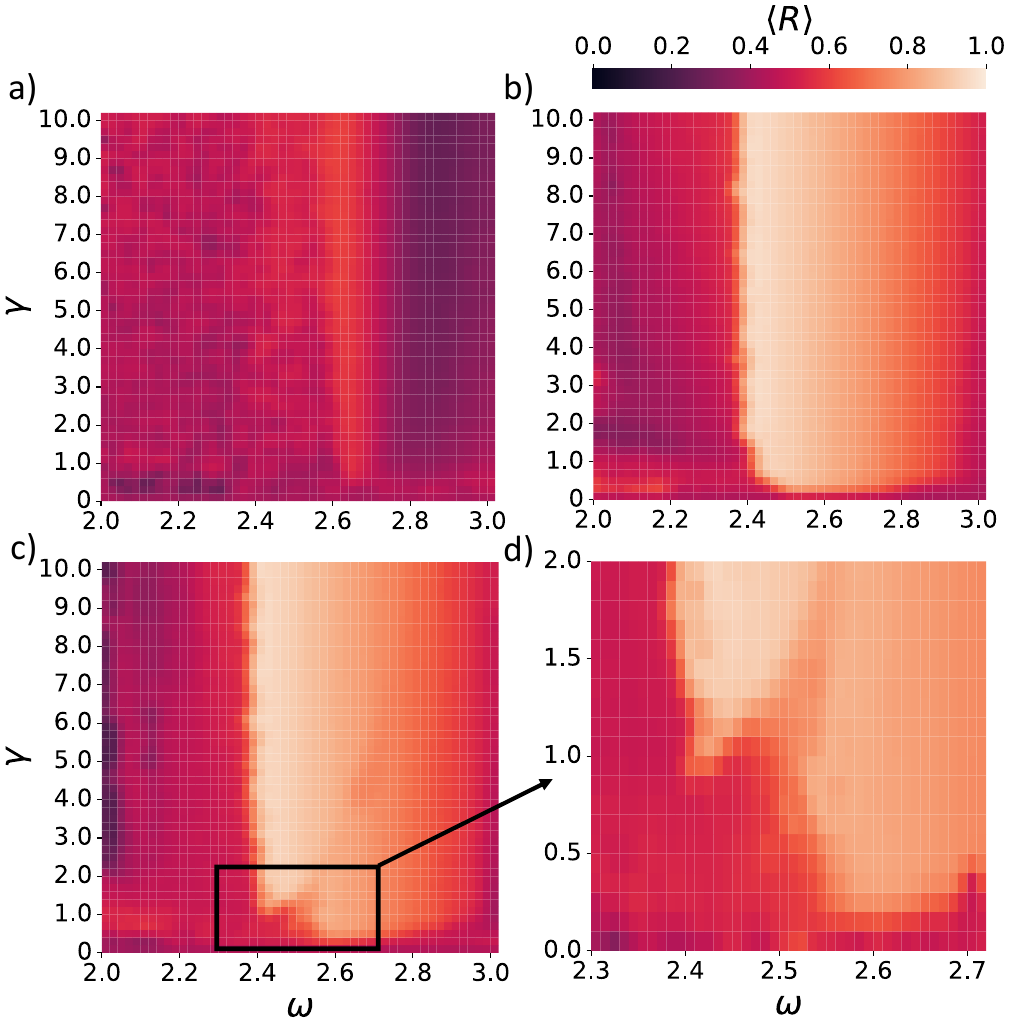}
    \caption{(color online) Synchronization regions in brain network with external stimulus: $\langle R \rangle$ in the parameter plane of the frequency $\omega$ and the input strength $\gamma$ of the external stimulus for cortical input regions (a) $k=14$, $I=1$, (b) $k=34$, $I=45$, and (c) $k=41$, $I=15$. Panel (d) shows a blow-up of (c). 
    Other parameters: $\sigma=0.7$, $\varsigma=0.15$, $\epsilon = 0.05$, $a=0.5$, and $\phi = \frac{\pi}{2} - 0.1$.}
    \label{fig.2}
\end{figure}

\begin{figure}
\centering
\includegraphics[width = \linewidth]{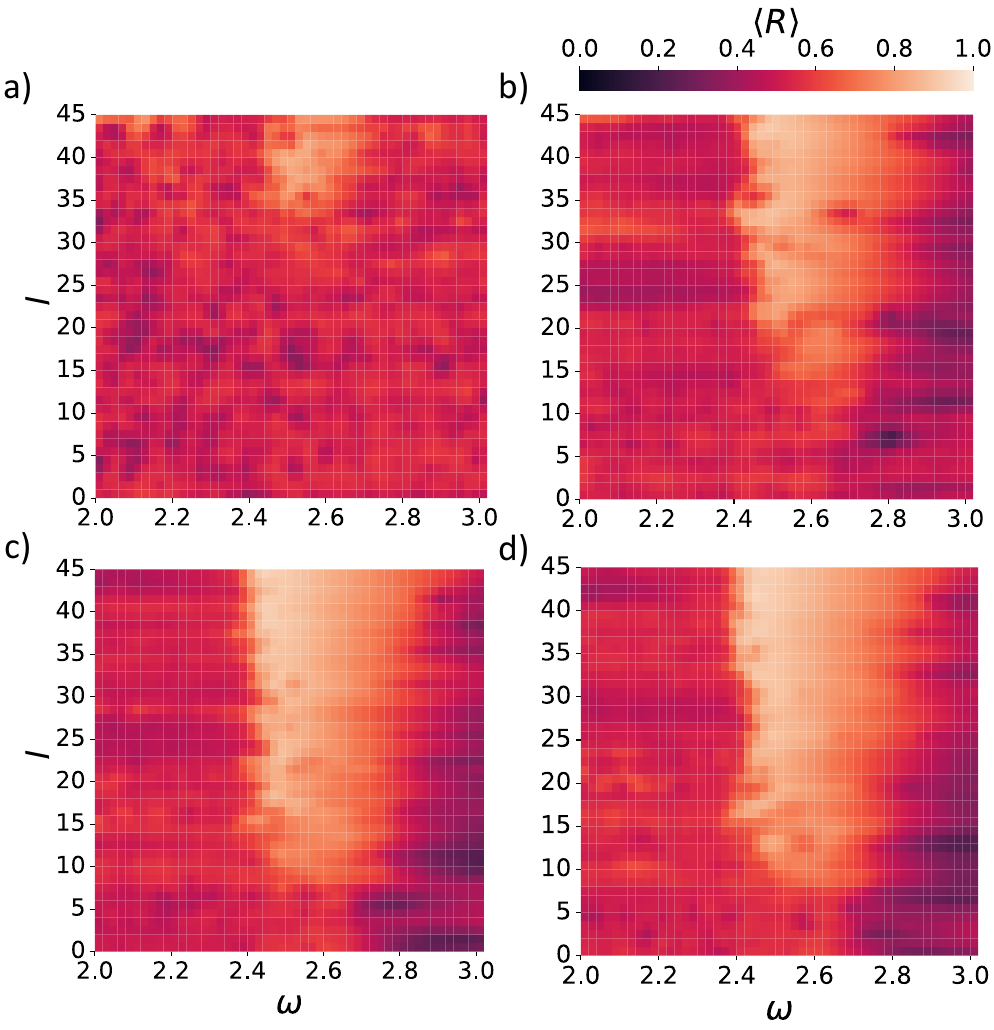}
    \caption{(color online) Same as in Fig.~\ref{fig.2}  in the parameter plane of the frequency $\omega$ of the external stimulus and its cortical input region $I$  (see Table \ref{tab:AAL}) for input strengths (a) $\gamma=0.11$, (b) $\gamma=1.1$, (c) $\gamma=7.0$, and (d) $\gamma=11.0$. Other parameters as in Fig.\,\ref{fig.2}.}
    \label{fig.3}
\end{figure}

In this section, we will investigate the role of the auditory cortex in the collective dynamics of the human brain. For this purpose, we feed a periodic external input into specific areas of our neural network, using the regions in pairs as described in the AAL atlas (Table \ref{tab:AAL}). Depending on the selected cortical regions $I$, a different influence on the degree of synchronization of the overall network can be observed as shown for three different input regions in Fig.\,\ref{fig.2}. The light colored regions in Fig.\,\ref{fig.2} indicate synchronized dynamics, whereas the darker colors indicate desynchronized dynamics. For $k=14$, $I=1$ in Fig.\,\ref{fig.2}a, there is a slightly light colored stripe around the natural uncoupled frequency $\omega=2.6$ ($\langle R \rangle \approx 0.7$). For $k=34$, $I=45$ in Fig.\,\ref{fig.2}b, there is a pronounced light colored synchronization region ($\langle R \rangle \approx 1$) starting at $\omega=2.4$. For $k=41$, $I=15$ in Fig.\,\ref{fig.2}c, a triangular  synchronization tongue splits off from the bottom left of the broad synchronization region, starting at $\omega=2.4$. This is shown in more detail in the close-up in Fig.\,\ref{fig.2}d. 
To further elaborate the different influence of different input regions, the global order parameter $\langle R \rangle$ is shown in Fig.\,\ref{fig.3} in dependence of the frequency $\omega$ of the external stimulus and its cortical input region $I$ for four values of the input strength $\gamma$. In the case of the auditory cortex ($I=15$), a distinct influence on the neuronal network can be observed, which is not so pronounced for other regions. There is a threshold for global synchronization at input frequencies of $\omega = 2.4$.
Even for very small input strengths $\gamma=0.11$ (see Fig.\,\ref{fig.3}a), synchronization of the entire network can be achieved for certain input regions $I$. On the other hand, even for very large input strengths $\gamma=7.0$ (see Fig.\,\ref{fig.3}c), some input regions never induce synchronization of the entire system. For better visibility, the input region indices $I ={\cal P}(1,...,N)$ (modulo 45) are permuted according to their synchronizability by re-arranging them in ascending order of the row sum of the temporal mean of the Kuramoto order parameters $\langle R \rangle$ averaged over all 4 panels of Fig.~\ref{fig.3}. Increasing the input strength further to $\gamma=11.0$ (see Fig.\,\ref{fig.3}d) does not lead to any quantitative change of the synchronization tongue.

Interestingly, the auditory cortex $I = 15$ (relabeled, marked yellow in Table \ref{tab:AAL}) is an input region that allows for the synchronization of the entire system at sufficiently large $\gamma$, but not at very small $\gamma$ (Fig.~\ref{fig.3}a). In this sense it plays an intermediate role between cortical areas which can easily synchronize the brain (marked pink in Table \ref{tab:AAL}) and those which never synchronize (marked blue in Table \ref{tab:AAL}).

\begin{table}[tbp!]
\centering
\resizebox{\linewidth}{!}{
\begin{tabular}{|c|c|c|c|} 
   \hline
   $k$: L/R & $I$ & Region &Lobe \\
   \hline
   \hline
   1/46  & 35 & Precentral &  Central region\\
   2/47  & 18 &Frontal Sup &  Frontal lobe\\
   \rowcolor{cyan}3/48  & 7 & Frontal Sup Orb & Frontal lobe\\
   4/49  & 31 &Frontal Mid &  Frontal lobe\\
   \rowcolor{cyan}5/50  & 8 & Frontal Mid Orb & Frontal lobe\\
   6/51  & 34 &Frontal Inf Oper &Frontal lobe \\
   7/52  & 29 &Frontal Inf Tri  &  Frontal lobe\\
   \rowcolor{cyan}8/53  & 3 &Frontal Inf Orb &  Frontal lobe\\
   9/54  & 11 &Rolandic Oper &  Central Region \\
   10/55 & 30 &Supp Motor Area &  Frontal lobe\\
   \rowcolor{cyan}11/56 & 5 &Olfactory & Frontal lobe\\
   12/57 & 33 &Frontal Sup Medial &  Frontal lobe\\
   13/58 & 21 &Frontal Med Orb  &  Frontal lobe\\
   \rowcolor{cyan}14/59 & 1 & Rectus & Frontal lobe\\
   15/60 & 22 &Insula  & Insula\\
   16/61 & 20 &Cingulum Ant  & Limbic lobe\\
   17/62 & 25 &Cingulum Mid & Limbic lobe\\
   18/63 & 26 &Cingulum Post & Limbic lobe\\
   19/64 & 13 &Hippocampus  & Limbic lobe \\
   20/65 & 16 &ParaHippocampal & Limbic lobe\\
   \rowcolor{cyan}21/66 & 2 &Amygdala & Sub cort gray nuc\\
   \rowcolor{pink}22/67 & 39 &Calcarine  & Occipital lobe\\
   \rowcolor{pink}23/68 & 43 &Cuneus  & Occipital lobe\\
   \rowcolor{pink}24/69 & 37 &Lingual  & Occipital lobe\\
   \rowcolor{pink}25/70 & 42 &Occipital Sup & Occipital lobe \\
   \rowcolor{pink}26/71 & 38 &Occipital Mid  & Occipital lobe\\
   27/72 & 32 &Occipital Inf  & Occipital lobe \\
   28/73 & 23 &Fusiform   & Occipital lobe\\
   \rowcolor{pink}29/74 & 41 &Postcentral & Central region\\
   30/75 & 28 & Parietal Sup & Parietal lobe \\
   31/76 & 36 & Parietal Inf  & Parietal lobe\\
   \rowcolor{pink}32/77 & 44 &Supramarginal & Parietal lobe \\
   \rowcolor{pink}33/78 & 40 &Angular  & Parietal lobe\\
   \rowcolor{pink}34/79 & 45 &Precuneus & Parietal lobe\\
   35/80 & 14 &Paracentral Lobule & Frontal lobe\\
   \rowcolor{cyan}36/81 & 9 &Caudate & Sub cort gray nuc\\
   37/82 & 19 &Putamen &Sub cort gray nuc\\
   38/83 & 12 &Pallidum &  Sub cort gray nuc\\
   39/84 & 17 &Thalamus & Sub cort gray nuc\\
   40/85 & 10 &Heschl &  Temporal lobe\\
   \rowcolor{yellow}41/86 & 15 &Temporal Sup & Temporal lobe\\
   \rowcolor{cyan}42/87 & 4 &Temporal Pole Sup & Limbic lobe\\
   43/88 & 24 &Temporal Mid & Temporal lobe\\
   \rowcolor{cyan}44/89 & 6 &Temporal Pole Mid & Limbic lobe\\
   45/90 & 27 &Temporal Inf &Temporal lobe \\ \hline
\end{tabular}}

\caption{Cortical and subcortical regions, according to the Automated Anatomical Labeling atlas (AAL) \cite{TZO02}. $I$ denotes the relabeled index of the regions ordered according to increasing influence on global synchronization. A small input signal ($\gamma = 0.11$, see Fig.\,\ref{fig.3}a) to the pink shaded regions has a big impact on the synchronization of the whole system ($\langle R \rangle > 0.8)$, whereas even strong inputs ($\gamma = 11.0$, see Fig.\,\ref{fig.3}d) to the blue shaded regions have no impact on synchronization ($\langle R \rangle < 0.5)$. The brain areas $k=41,86$ (yellow shaded, $I=15$) are associated with the auditory cortex.   
}
\label{tab:AAL}
\end{table}

\begin{figure}
\centering
\includegraphics[width = 1.0\linewidth]{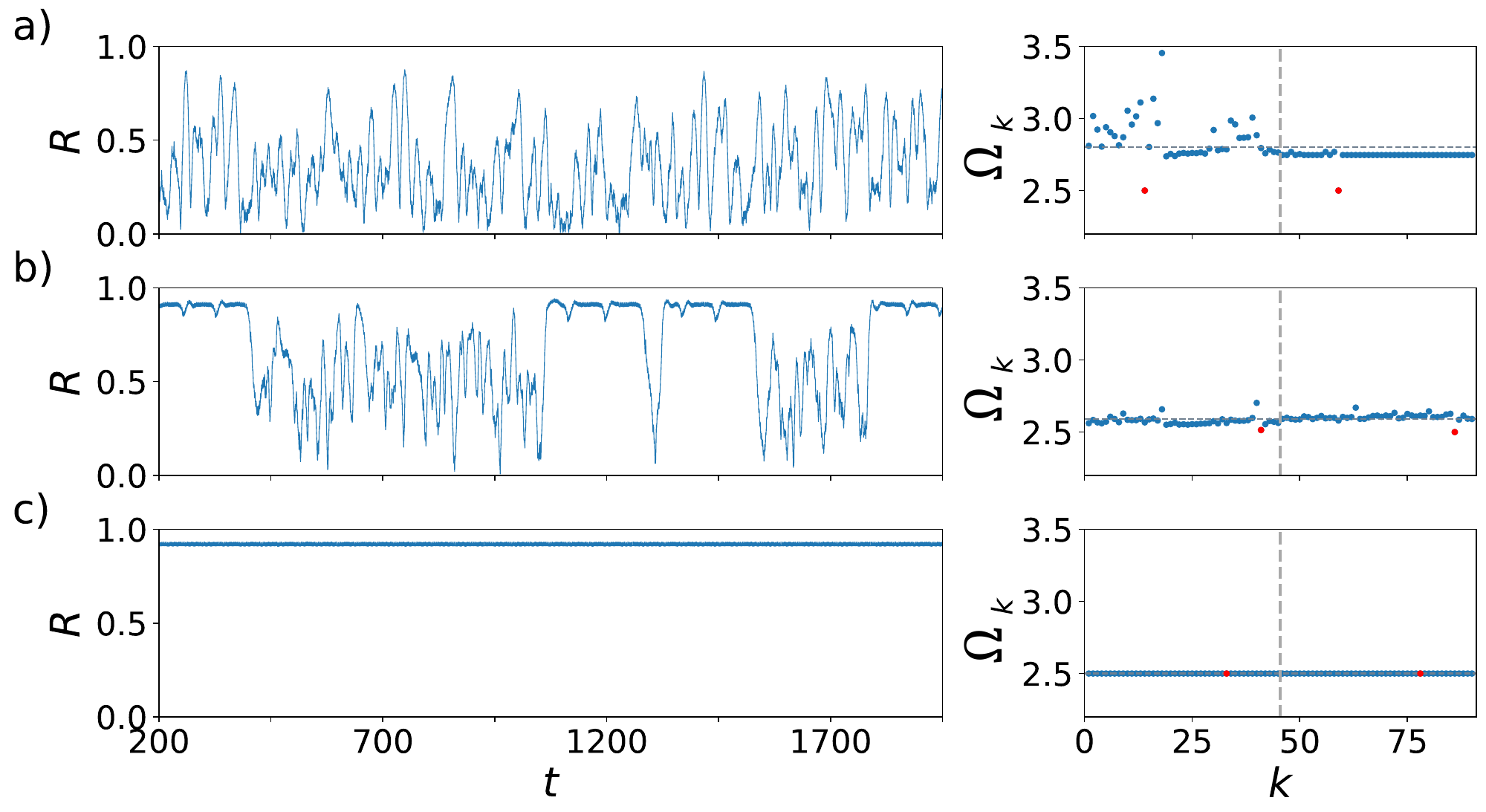}
    \caption{(color online) Dynamical scenarios: 
    $R(t)$ (left column) and mean phase velocities $\Omega_k$ (right column) for frequency $\omega=2.5$ of the external stimulus and fixed amplitude $\gamma = 1.1$ applied to different input regions (a) $k=14$, $I=1$, (b) $k=41$, $I=15$, (c) $k=34$, $I=45$. The vertical dashed line in the right column separates the left and right hemisphere; the horizontal grey dotted line indicates the temporal average of the mean-field frequency $\Omega$. The red dots mark the nodes of (a) the rectus $k=14,59$, (b) the auditory cortical regions ($k=41,86$), and (c) the precuneus $k=34,79$. Other parameters as in Fig.\,\ref{fig.3}.}
    \label{fig.4}
\end{figure}

Figure \ref{fig.4} shows details of the synchronization dynamics for the three different groups of input regions in terms of the time evolution of $R$ (left) and the mean phase velocities (right). Fixing the input amplitude $\gamma$ and frequency $\omega$ beyond the threshold of the synchronization tongue, we find three different scenarios: In Fig.\,\ref{fig.4}a, the time evolution of the Kuramoto order parameter is similar to the system behavior without external stimulus, i.e., it exhibits large time variations and no global synchronization. The mean phase velocities of all nodes, in particular in the left hemisphere, show no frequency synchrony, except for some local areas, and stronger frequency synchrony in the right hemisphere, which displays a remarkable asymmetry. The horizontal gray dotted line shows the time average of the collective mean field frequency $\Omega$, which in the right hemisphere agrees well with the individual mean phase velocities $\Omega_k$, however it is clearly distinct from the frequency of the two stimulated areas (red dots). Fig.\,\ref{fig.4}b shows input to the auditory cortex, leading to episodes of strong global synchrony interrupted by intervals of desynchronization (left panel), and overall mean phase velocities close to the collective mean frequency $\Omega$ (right panel). Fig.\,\ref{fig.4}c shows constantly high global synchronization equal to the input frequency. Thus the auditory cortex plays an intermediate role between brain areas which cannot synchronize the whole system and others which do so strongly in a very stable way. 

\begin{figure}
\centering
\includegraphics[width=0.8\linewidth]{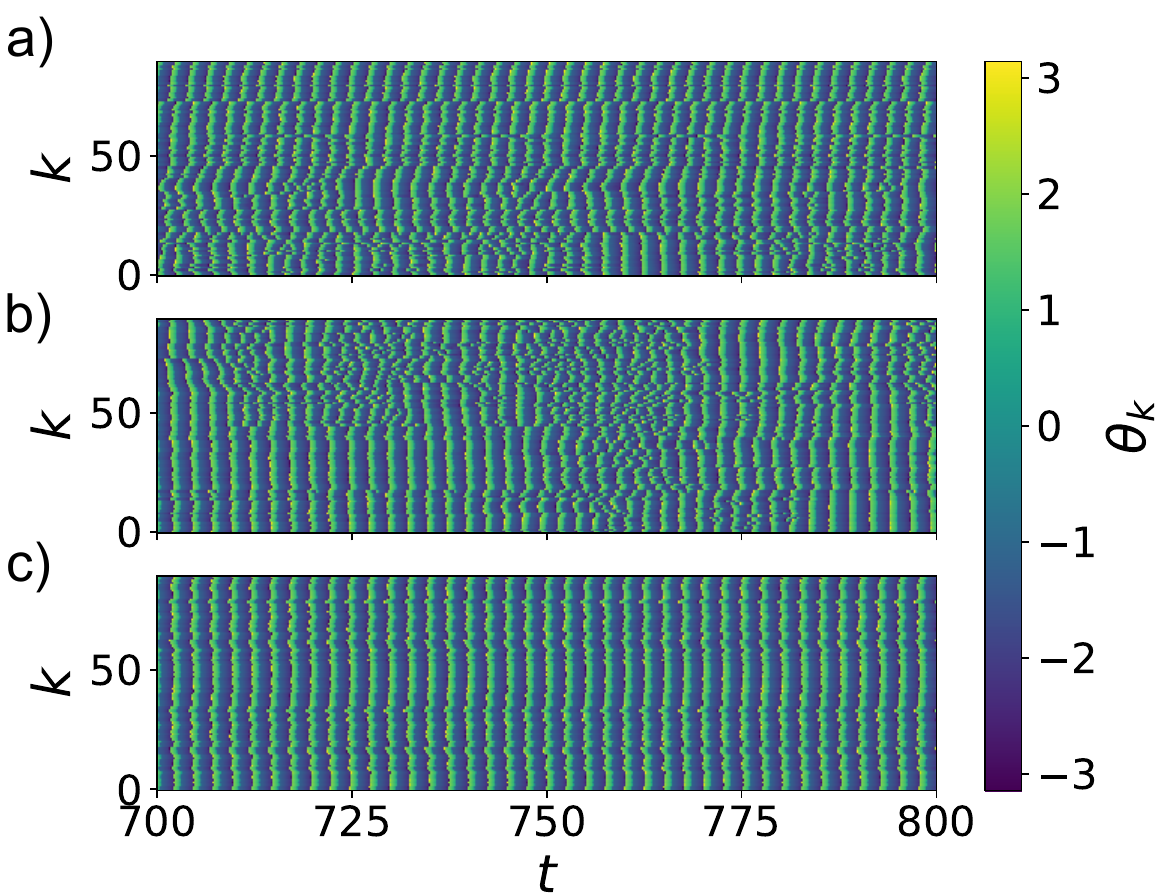}
\caption{(color online) Partially synchronized dynamics: space-time plot of the phase variable $\theta_k=\arctan(u_k,v_k)$ for frequency $\omega=2.5$ of the external stimulus and fixed amplitude $\gamma = 1.1$ applied to different input regions (a) $I=1$, (b) $I=15$, (c) $I=45$, corresponding to Fig.\,\ref{fig.4}a,b,c, respectively. Other parameters are as in Fig.\,\ref{fig.3}.}
\label{fig.5}
\end{figure}

For further insight into the local synchronization, Fig.\,\ref{fig.5} shows the space-time representation of the phase variable $\theta_k$ for the corresponding parameters in Fig.\,\ref{fig.4}. Figure \ref{fig.5}a shows quite irregular dynamics independent of the cortical input, while in Panel (c) strong global synchronization is induced by the cortical input. In Panel (b) the alternating synchronization and desynchronization dynamics can be clearly seen. 
 This delicate balance of synchrony and asynchrony for auditory stimulation is probably typical for the critical state and the high sensitivity of the brain operating between different dynamic regimes. 

\section{Conclusions and future challenges}

The interplay of synchronization and cortical input is essential in understanding the perception of music in the brain. 
Using simple oscillator models on the nodes of the network in combination with complex brain network structures and input into different cortical areas is one step towards insight into the 
network response generated by the structural connectivities of the stimulated node for different locations of input. In particular, we observe distinctly different synchronization dynamics if the periodic input is fed into nodes which are not part of the auditory cortex.
We have employed empirical structural network connectivities obtained from diffusion-weighted magnetic resonance imaging of humans together with paradigmatic dynamics of the FitzHugh-Nagumo model. This yields significant scenarios of partial synchronization of the brain, and supports the general observation that the brain is operating in a critical state at the edge of different dynamical regimes~\cite{MAS15a,WIL19,SHI22}.

Specifically, we 
found episodes of higher synchrony that depend on the frequency, amplitude, and location of the injection. We have shown that two frequencies play an essential role in synchronization, namely the natural frequency of the uncoupled oscillator and the collective frequency of the coupled system. The degree of synchronization gradually increases as the frequency of input approaches the frequency of the uncoupled oscillator or multiples thereof.  On the other hand, the choice of the cortical input region is equally important for the synchronizability of the neuronal network. The auditory cortex plays a prominent role in this respect because it induces variability of synchrony.  In this context, a sharp difference between the impact of stimulations targeted at a specific module of the network and stimulations distributed over the network has also been noted in ~\cite{FRA18b}. It would be interesting to elaborate the effect of input into various sensory cortices, for which different brain atlases need to be related~\cite{BOH09,EVA12}.

These results are consistent with experiments \cite{HAR20a, SAW22} showing that music evokes some degree of synchrony in the human brain. In these works it has been shown that listening to music can have a remarkable effect on brain dynamics, inducing in particular a periodic alternation between synchronization and desynchronization. Indeed, such an alternation reflects the variability of the brain; it can be considered as a critical state between a fully synchronized and a fully desynchronized state. It is well known that in a critical state, the brain moves on the edge of different dynamical regimes and exhibits hysteresis and avalanche phenomena, as observed in critical phenomena and phase transitions \cite{RIB10,STE10a,KIM18,SCA18,CON23}. By tuning the parameters, we have found intriguing dynamical scenarios during the transition to synchronization and observed the intermittent change between high and low degrees of synchronization \cite{SAW21a}. In summary, simulating an external sound source connected to the brain enables synchronization dynamics that can be used to model the effect of music on the human brain. 

In a more detailed study of the perception of music in brain network dynamics using a cochlea model for the auditory input~\cite{SAW22}, the influence of real music in a simulated network of FitzHugh-Nagumo oscillators with empirical structural connectivity has been investigated, and compared to measured EEG data. An increase of coherence between the global dynamics and the input signal induced by a specific music song has been found. It has been shown that the level of coherence depends sensitively on the frequency band. The simulation  results have been compared with experimental data, which describe global neural synchronization between different brain regions in the gamma-band range and its increase just before transitions between different parts of the musical form (musical high-level events). Synchronization increases before musical large-scale form boundaries, and decreases afterwards, therefore this represents musical large-scale form perception.  

\acknowledgments
\noindent This work was supported by the Deutsche Forschungsgemeinschaft (DFG) - Projects No. 429685422 and 440145547. We are grateful to Rolf Bader and Lenz Hartmann for stimulating discussion and collaboration.


\end{document}